\documentclass[9pt,preprint2,natbib209]{aastex61}

\usepackage{graphicx}
\usepackage{dcolumn}
\usepackage{bm}

\def\apjs{ApJS}

\def\apjl{ApJL}

\def\aap{A\&A}

\begin{document}


\title{Possible Formation Scenarios of ZTF J153932.16+502738.8-A Gravitational Source Close to the Peak of LISA¡¯s Sensitivity}

\email{guolianglv@xao.ac.cn}
\author{Guoliang L\"{u}}
\affil{School of Physical Science and Technology,
Xinjiang University, Urumqi, 830046, China}
\affil{Center for Theoretical Physics,
Xinjiang University, Urumqi, 830046, China}

\author{Chunhua Zhu}
\affil{School of Physical Science and Technology,
Xinjiang University, Urumqi, 830046, China}
\affil{Center for Theoretical Physics,
Xinjiang University, Urumqi, 830046, China}

\author{Zhaojun Wang}
\affil{School of Physical Science and Technology,
Xinjiang University, Urumqi, 830046, China}
\affil{Center for Theoretical Physics,
Xinjiang University, Urumqi, 830046, China}

\author{Helei Liu}
\affil{School of Physical Science and Technology,
Xinjiang University, Urumqi, 830046, China}
\affil{Center for Theoretical Physics,
Xinjiang University, Urumqi, 830046, China}

\author{Lin Li}
\affil{School of Physical Science and Technology,
Xinjiang University, Urumqi, 830046, China}
\affil{Center for Theoretical Physics,
Xinjiang University, Urumqi, 830046, China}

\author{Dian Xie}
\affil{School of Physical Science and Technology,
Xinjiang University, Urumqi, 830046, China}

\author{Jinzhong Liu}
\affil{National Astronomical Observatories / Xinjiang Observatory,
the Chinese Academy of Sciences, Urumqi, 830011, China}


\date{\today}




\begin{abstract}
ZTF J153932.16+502738.8 (ZTFJ1539) is an eclipsing double-white-dwarf system with an orbital period of 6.91 minutes,
and is a significant source of LISA detecting gravitational wave. However, the massive white dwarf (WD) with mass of about 0.61 M$_\odot$
has a high effective temperature (48900 K), and the lower mass WD with mass of about 0.21 M$_{\odot}$ has a low effective temperature($<$10000 K).
It is challenging the popular theory of binary evolution. We investigate the formation of ZTFJ1539 via nova and Algol scenarios.
Assuming that the massive WD in ZTFJ1539 just experiences a thermalnuclear runaway,
nova scenario can explain the effective temperatures of two WDs in ZTFJ1539. However,
in order to enlarging a semi-detached orbit of about 4---5 minutes to a detached orbit of about 7 minutes,
nova scenario needs a much high kick velocity of about 200 km s$^{-1}$ during nova eruption. The high kick velocity can result in
high eccentricity of about 0.2---0.6.
Algol scenario can also produce ZTFJ1539 if we take a high efficient parameter for ejecting common envelope
and enhance the mass-loss rate via stellar wind trigger by
tidal effect.
\end{abstract}

\section{Introduction}
The gravitational radiation can be produced by these binaries with short orbital periods.
The LIGO/Virgo detectors have successfully observed the gravitational waves (GWs) emitted in the kHz regime
by several mergers of double black holes and a merger of a double neutron stars\citep{Abbott2016,Abbott2017,Abbott2018}.
In fact, comparing with double black holes and neutron stars, double white dwarfs (DWDs) are dominated\citep{Nelemans2001a}.
In the Milk Way, \cite{Nelemans2001b} estimated that there are about $10^7$ DWDs emitting gravitational waves in the mHz band.
These GWs can be detected by the Laser Interferometer Space Antenna (LISA) which can
observe the GWs with a band of about 0.1 and 1000 mHz\citep{Amaro2017}.
\cite{Cornish2017} estimated that LISA can find about $10^4$ DWDs.

Recently, based on the updated distances from {\it Gaia} Data Release 2, \cite{Kupfer2018}
gave 13 binary systems which can emit GWs strong enough to be detectable by LISA.
Very recently, \cite{Burdge2019} observed an eclipsing DWD ZTF J153932.16+502738.8 (ZTFJ1539)
with an orbital period of 6.91 minutes, which is the shortest in detached DWDs.
The GW emitted by this DWD is closed to the most sensitive LISA's band,
and LISA can detect it within a week\citep{Burdge2019}. Therefore, ZTFJ1539 is
a significant source of LISA detecting GW\citep{Littenberg2019}

For the theory of binary evolution, scenario is very interesting.
The primary and secondary WDs have masses of about 0.61 and 0.21 M$_\odot$\citep{Burdge2019}, respectively.
In this paper, the WD with larger mass is called as the primary WD, while another is called as the secondary WD.
Usually, in DWDs, the primary WD first forms, and its cooling timescale is shorter than that of the secondary WD.
Surprisedly, the effective temperature of the primary WD is about 48900$\pm$900 K, which is
higher than the temperature ($<10000$ K) of the secondary WD\citep{Burdge2019}.
\cite{Burdge2019} gave two possible explanations. One is tidal heating. Due to extremely short orbital period,
tidal distortion dissipates the tidal energy in the WD, which results in heating and spinning up WD\citep{Fuller2013}.
Another is recent accretion which results in a nova eruption.
Having considered that the temperature of WD heated by the tidal distortion through more
realistic calculations is closed to 25000 K, \cite{Burdge2019} suggested the latter scenario better.

Theoretically, in DWDs, the accretion should occur in semi-detached systems, and result in X-ray emission.
However, ZTFJ1539 is a detached system and is no detectable X-ray emission\citep{Burdge2019}.
Therefore, there should be a mechanism for a transition from semi-detached system to detached system
if ZTFJ1539 undergoes nova eruption.
In addition, it is possible that the secondary WD may form earlier than the primary WD if binary experiences
mass transfer to become Algol system\citep[e. g.,][]{Tout2010,Rensbergen2011}.

In this paper, we investigate the possibility for ZTFJ1539 produced by nova eruption with a kick velocity or
Algol systems. In \S 2 section, we simulate the nova eruptions in DWDs, and discuss the possibility of forming
ZTFJ1539 via nova scenario. The Algol scenario producing ZTFJ1539 is discussed in \S 3. The main conclusions
appear in \S 4.

\section{Nova Scenario}
Nova eruption usually appears in binary system in which
a thermal nuclear runaway occurs on the surface of WD accreting
material from its companion. In general, most of companions in nova binaries are normal stars (main sequence or giant stars),
while about 60 nova binaries have hydrogen-deficient companions\citep[e. g.,][]{Gallagher1978,Belczynski2000,Lu2006,Ramsay2018}.
The former is called as cataclysmic variables (CV),
and the latter is called as ultra-compact cataclysmic variables (AM CVn).
However, ZTFJ1539 is a DWD binary. It is composed of a CO WD with mass of about 0.61 M$_\odot$ and a He WD
with mass of about 0.21 M$_\odot$\citep{Burdge2019}.
If the He WD fills its Roche lobe, mass transfer occurs.
\cite{Marsh2004} investigated the stability of mass transfer between DWDs. 
If mass transfer is stable, ZTFJ1539 will undergo a nova eruption,
and evolve into an AM CVn\citep{Burdge2019}. If the transfer is not stable, it could evolve into
a R Coronae Borealis Stars\citep[e. g.,][]{Clayton2012,Han1998}.
In order to simulate the nova eruption in ZTFJ1539, we must construct the models of He and CO WD.

\subsection{Nova model}
\begin{figure*}
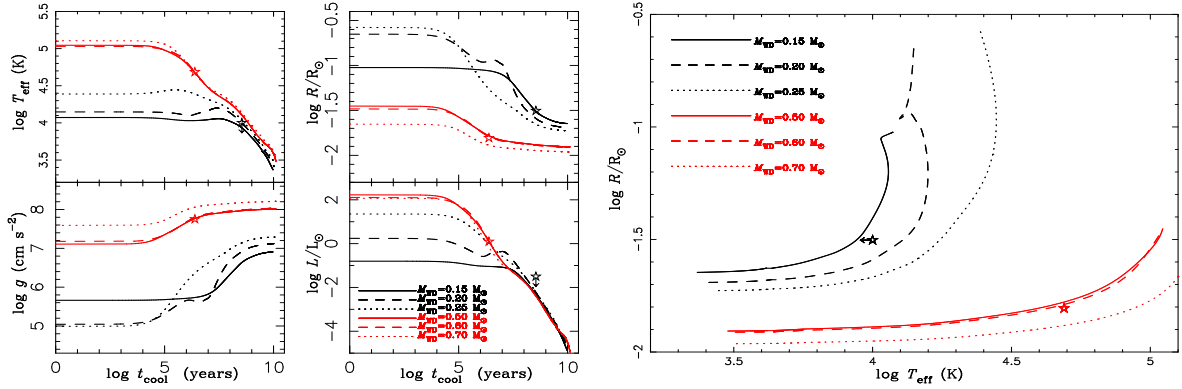

\begin{tabular}{cc}
\includegraphics[totalheight=3.in,width=2.in,angle=-90]{wdcl.ps}&
\includegraphics[totalheight=3.in,width=2.in,angle=-90]{rate.ps}\\
\end{tabular}
\caption{The evolutions of the effective temperature, radius, acceleration gravity and luminosity during
WD cooling phase are given in the left panels. The effective temperature vs. the radius is showed in the right
panel. The black and red lines represent He and CO WDs, respectively. The line styles represent WDs with different
masses. The black and red stars give the positions of He and CO WDs in ZTFJ1539. The observational data come from \cite{Burdge2019}.  }
\label{fig:wdcl}
\end{figure*}

Modules for Experiments in Stellar Evolution (MESA,
[rev. 11108]; \cite{Paxton2011,Paxton2013,Paxton2015,Paxton2018}) offers a test$\_$suite for
producing He WD and CO WD.
Using MESA, we create CO WDs with masses of 0.5, 0.6 and 0.7 M$_\odot$, and He WDs with masses of 0.15, 0.20 and 0.25 M$_\odot$, respectively.
Figure \ref{fig:wdcl} shows the evolutions of the effective temperature, radius, gravity acceleration and luminosity during
WD cooling phase.
The cooling age of a nascent WD with mass of 0.6  or 0.2 M$_\odot$ which evolves to the state of the primary or the secondary in
ZTFJ1539 is respectively about 2.4 M or 300 M years, which is consistent with the estimate of \cite{Burdge2019}.
Based on Figure \ref{fig:wdcl}, the primary or the secondary can be covered by single-star model.
However, ZTFJ1539 is a DWD.
According to popular binary theory, the WD with mass higher in DWD should be cooler because
it is older and its cooling timescale is shorter.
\cite{Burdge2019} considered that the primary in ZTFJ1539 just experienced a nova eruption.
Its primary and secondary are CO WD and He WD, respectively.
It means that the nova is triggered by the helium-rich material accreted by the primary
from the secondary. Very recently, using the FUN evolutionary code combined with
a large nuclear network, \cite{Piersanti2019} simulated the nova eruptions produced
by helium-accreting WD in AM CVn stars.

In this work, following \cite{Denissenkov2013,Denissenkov2014,Zhu2019},
we use test$\_$suit case nova in MESA to simulate nova eruption in ZTFJ1539.
Based on the observations of Keck I, \cite{Burdge2019} found that
narrower hydrogen emission lines apparently come from the cooler secondary,
and the irradiated surface of the secondary had some weak neutral helium emission lines.
It is difficult to determine the chemical compositions of the secondary.
In our model, we simulate nova eruptions trigged by hydrogen-rich or hydrogen-deficient
accreted materials. The chemical compositions of the former are $X({\rm H})=0.7$, $X({\rm He})=0.28$ and $Z=0.02$,
they for the latter are $X({\rm H})=0.0$, $X({\rm He})=0.98$ and $Z=0.02$. The abundances of metal elements are
similar with those of the Sun.

\subsection{Discussions}
\begin{figure*}
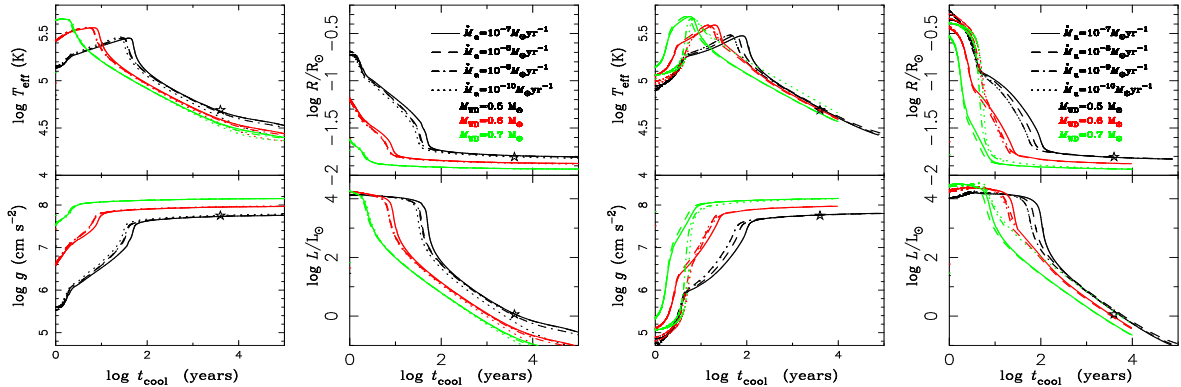

\begin{tabular}{cc}
\includegraphics[totalheight=3.in,width=2.in,angle=-90]{nvclold.ps}&
\includegraphics[totalheight=3.in,width=2.in,angle=-90]{nvcl.ps}\\
\end{tabular}
\caption{The evolutions of the effective temperature, radius, gravitational acceleration and luminosity
after novae reach the maximum luminosity. The left four panels are for nova eruptions trigged by hydrogen-rich
accreted materials, and the right four panels are for hydrogen-deficient accreted materials, respectively.
 The black, red and green lines represent the nova models with CO WD's
masses of 0.5, 0.6  and 0.7 M$_\odot$, respectively. The solid, dashed, dot-dashed and dotted lines represent
the nova models with the mass-accretion rates of $10^{-7}$, $10^{-8}$, $10^{-9}$ and $10^{-10}$ M$_\odot$yr$^{-1}$,
respectively. The black star gives the observational value of the primary in ZTFJ1539\citep{Burdge2019}. }
\label{fig:nvcl}
\end{figure*}

Figure \ref{fig:nvcl} gives the evolutions of the effective temperatures, the radii, the gravitational acceleration and the luminosities for
WDs after novae reach the maximum luminosity. Both of hydrogen-rich and hydrogen-deficient models give
similar results.
The primary of ZTFJ1539 can be explained by WD which has experienced a nova eruption
about 8000 yr ago. Compared with the radius and the gravitational acceleration of the primary,
the model of WD with 0.5 M$_\odot$ mass is better. However, both of them greatly depend on the hydrogen mass around
WD surface. In our simulation, we consider the mass loss when the luminosity of nova exceeds the Eddington luminosity,
and may underestimate the hydrogen mass. In addition, they also depend on the WD cooling which
is also affected by the convection, the radiative
transfer, crystallization, and so on \citep{Wood1992, Hansen1999,Liu2019}.

Usually, the nova in DWDs should occur in semi-detached systems. However,
ZTFJ1539 is a detached binary. \cite{Burdge2019} considered that the orbit
was widen because an amount of mass ($10^{-4}$M$_\odot$) was ejected during
nova eruption.
In order to discuss this effects,
considering that the nova eruption may be aspherically symmetric
and following the method of investigating the kick velocity of supernova in \cite{Brandt1995,Hurley2002},
we assume that the WD also receives a kick velocity when it ejects an amount of mass.
The kick velocity $v_{\rm k}$ abides by Maxwellian distribution
\begin{equation}
P(v_{\rm k})=\sqrt{\frac{2}{\pi}}\frac{v^2_{\rm k}}{\sigma^3_{\rm
k}}e^{-v^2_{\rm k}/2\sigma^2_{\rm k}}.
\end{equation}
In this work, $\sigma_{\rm k}$ is taken as 50 and 200 km s$^{-1}$, respectively.

By a method of synthesis population, we construct a sample of $10^7$ binary systems\citep[Details can be seen in][]{Lu2011,Lu2012,Lu2013,Zhu2013,Zhu2015}.
The initial mass function comes from \cite{Miller1979}, a constant mass-ratio distribution is considered \citep{Mazeh1992},
and the distribution of separations is taken as
\begin{equation}
\log a = 5X+1
\end{equation}
where $X$ is a random variable uniformly distributed in the range [0,1]
and $a$ is in $R_\odot$.

\begin{figure}
\begin{tabular}{cc}
\includegraphics[totalheight=3.in,width=2.5in,angle=-90]{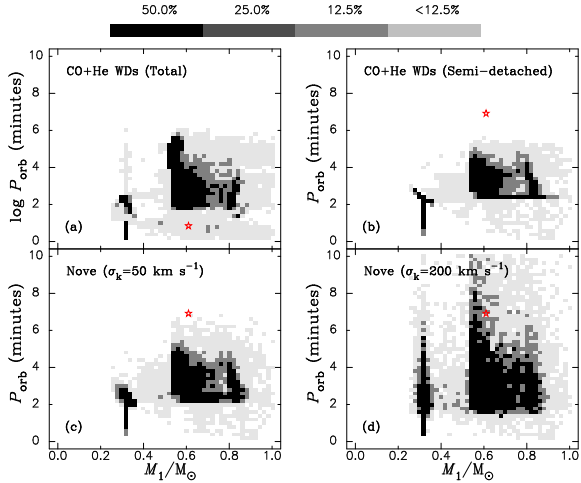}
\end{tabular}
\caption{The primary masses vs. the orbital periods of CO + He WD systems. The panel (a) represents
 the nascent CO + He WDs, and the panel (b) gives the semi-detached CO + He WDs. The panels (c) and (d)
 show the CO + He WDs after nova eruptions with kick velocity of $\sigma_{\rm k}$= 50 and 200 km s$^{-1}$,
 respectively. The red star gives the position of ZTFJ1539.}
\label{fig:kickmp}
\end{figure}

Using the rapid binary star evolution (BSE) code \citep{Hurley2002}, we obtain
about $3.4\times10^5$ CO + He WDs from $10^7$ binary systems.
The panel (a) of Figure \ref{fig:kickmp} gives the distribution of the primary masses and the orbital periods for nascent
CO + He WDs. Due to the gravitational release, the orbits of these systems shrink.
Within Hubble times, about 44\% of them evolve into semi-detached systems which are showed in
the panel (b) of Figure \ref{fig:kickmp}. Because the He WDs in our simulations have very thick hydrogen envelopes ($\sim 10^{-5}$M$_\odot$),
their orbital periods are shorter than about 6 minutes when the He WDs fill their Roche lobe.
If some He WDs have very thick hydrogen envelopes, and thus can fill their Roche lobes at longer periods.
When the WD with larger mass in the semi-detached systems accretes the matter from its companion,
the nova eruption occurs. \cite{Burdge2019} found that the orbit of ZTFJ1539 only widened slightly if mass of $10^{-4}$M$_\odot$ was ejected
during a nova eruption, and ZTFJ1539 would become a semi-detached system again in about 100 yr after a nova eruption.
However, as Figure \ref{fig:nvcl} shows, the primary of ZTFJ1539 may have evolved for about 8000 yr after a nova eruption.
In fact, based on the theoretical model, the mass ejected during a nova eruption is between about $10^{-3}$
and $10^{-8}$M$_\odot$\citep{Yaron2005,Zhu2019}. Compared with WD mass, the mass ejected is too small. It hardly results
in a large orbital widening.

\begin{figure}
\begin{tabular}{cc}
\includegraphics[totalheight=3.in,width=2.5in,angle=-90]{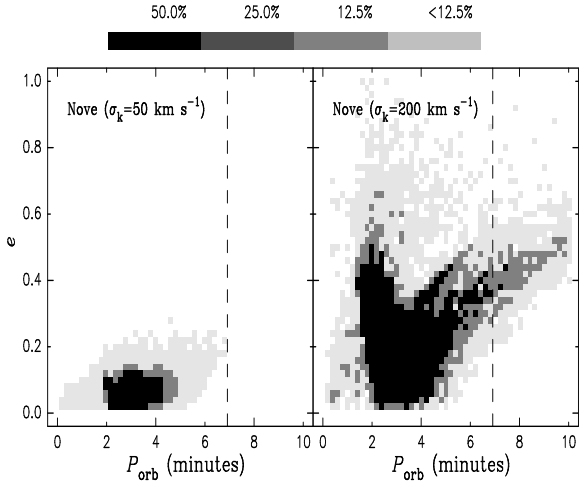}
\end{tabular}
\caption{The distributions of the orbital periods and eccentricities for
the CO + He WDs after nova eruptions with kick velocity of $\sigma_{\rm k}$= 50 and 200 km s$^{-1}$,
 respectively. The dashed line gives the orbital period of ZTFJ1539. }
\label{fig:kickep}
\end{figure}

In our work, we consider the effects of kick velocity. However, as the panel (c) of Figure \ref{fig:kickmp} shows,
the kick velocities with $\sigma_{\rm k}$ = 50 km s$^{-1}$ can not widen the orbits up to 6 minutes.
When $\sigma_{\rm k}$ increases to 200 km s$^{-1}$, ZTFJ1539 can be covered by our simulation.
Simultaneously, kick velocity can also result in an elliptical orbit. Figure \ref{fig:kickep} shows
the distributions of eccentricities for $\sigma_{\rm k}$ = 50  and 200 km s$^{-1}$, respectively.
Our results mean that the eccentricity of ZTFJ1539 should be between about 0.2 and 0.6 if the orbit of
ZTFJ1539 is broadened from about 5 minutes to 6.91 minutes by nova eruption.
Usually, a binary system with high eccentricity undergoes a large degree of
apsidal precession which would manifest itself in the eclipse timing residuals.
However, \cite{Burdge2019} believed that the eccentricity of ZTFJ1539 should be zero
because any sign of apsidal precession was not be detected.

In fact, \cite{Hobbs2005} analysed the proper motion of 233 pulsars, and suggested that
$\sigma_{\rm k}$ of core-collapse supernova (CCSN) is about 265 km s$^{-1}$. In general,
the energy released by CCSN is higher than $10^4$ times energy from nova eruption.
Therefore, $\sigma_{\rm k}$ of nova eruption should be very small.
In short, it is not a probability to explain the observational properties of ZTFJ1539
by experiencing a nova eruption.


\section{Algol Scenario}
Algol system is a binary where the less massive donor fills its Roche
lobe and transfers its matter to the more massive gainer which does not fill its Roche lobe,
and the former is the cooler, fainter and older star while the latter is still on the main sequence\citep{Peters2001}.
In the present paper, Algol scenario means that the progenitor of ZTFJ1539 undergoes the phase of Algol binaries so that
the less massive WD forms earlier than the more massive WD.

\subsection{Model of Binary Evolution}
It is well known that the interaction of binary systems produces many different types of astronomical phenomena.
However, some involved physical processes are still open, such as common envelope evolution (CEE).
The importance of CEE in binary evolution has been discussed by many literatures\citep[e. g.,][]{Ivanova2013}.
In general, the formation of DWD system experiences CEE once or twice\citep[e. g.,][]{Postnov2014}.
Unfortunately, it is extremely challenging to simulate real CEE via both computation and analytic treatment.
In the last section, using BSE code, we obtain about $3.4\times10^5$ CO + He WDs via a method of population synthesis.
In BSE code, the treatment of CEE is determined by the initial binding energy of the envelope and the initial orbital
energy of two cores\citep[Details can be seen in][]{Hurley2002}. CEE is affected by two parameters:
$\alpha_{\rm CE}$ and $\lambda$. The former ($\alpha_{\rm CE}$ ) is the efficiency of the orbital energy
used to expel the envelope\citep{Paczynski1976}, and the latter ($\lambda$) is related to the structure of the donor\citep{Kool1990,Xu2010}.
In fact, the outcome of CEE in BSE code depends on the product of $\alpha_{\rm CE}$ and $\lambda$\citep{Hurley2002,Lu2006,Lu2012}.

\begin{figure}
\includegraphics[totalheight=3.in,width=2.5in,angle=-90]{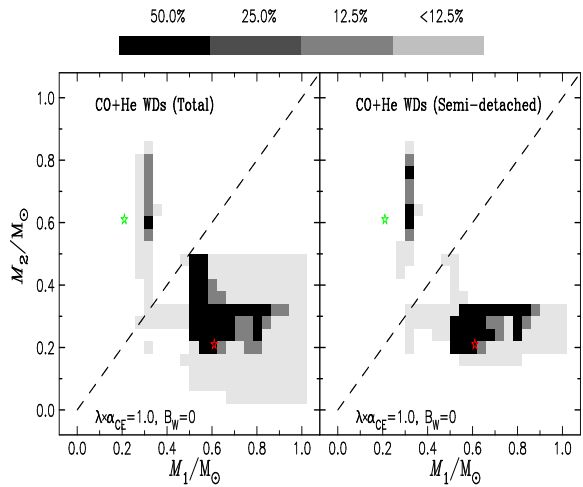}
\caption{Distribution of WDs' masses when He + CO WD systems form. The left panel represents
all He + CO WD systems, while the right panel is for these He + CO WD systems which can merge within
the Hubble time. Red star represents ZTFJ1539 where $M_1$ is the primary and $M_2$ is the secondary, that is, $M_1=0.61$M$_\odot$
and $M_2=0.21$M$_\odot$. However, blue star represents ZTFJ1539 where $M_1=0.21$M$_\odot$
and $M_2=0.61$M$_\odot$. Dashed line means a position where $M_1=M_2$. The He + CO WD systems locating
the left-top zone is produced via Algol scenario. }
\label{fig:m1m2c1}
\end{figure}

In the previous section, we use combining parameter $\lambda\times\alpha_{\rm CE}=1.0$ to calculate binary evolutions.
Figure \ref{fig:m1m2c1} shows the distribution of WDs' masses when He + CO WD systems form.
Obviously, He + CO WD systems can be produced via Algol scenario. These systems is showed in the
left-top zone of Figure \ref{fig:m1m2c1}. Unfortunately, the He WD's masses of these systems are
higher than about 0.25 M$_\odot$, while the secondary's mass of ZTFJ1539 is only 0.21 M$_\odot$.
However, the products of binary systems are affected by many uncertainty input parameters.

\begin{figure*}
\begin{tabular}{cc}
\includegraphics[totalheight=4.5in,width=3.in,angle=0]{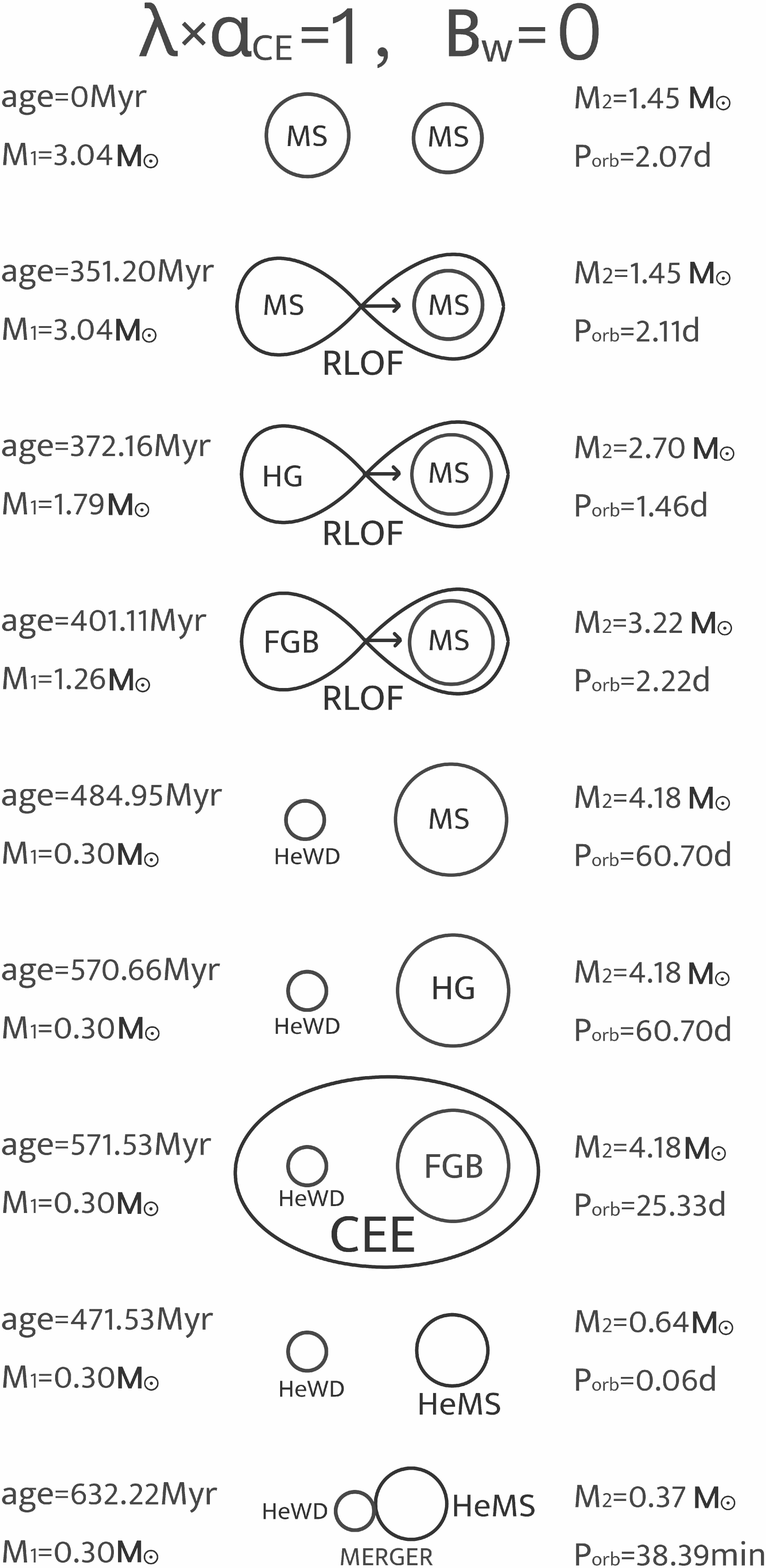}&
\includegraphics[totalheight=4.5in,width=3.in,angle=0]{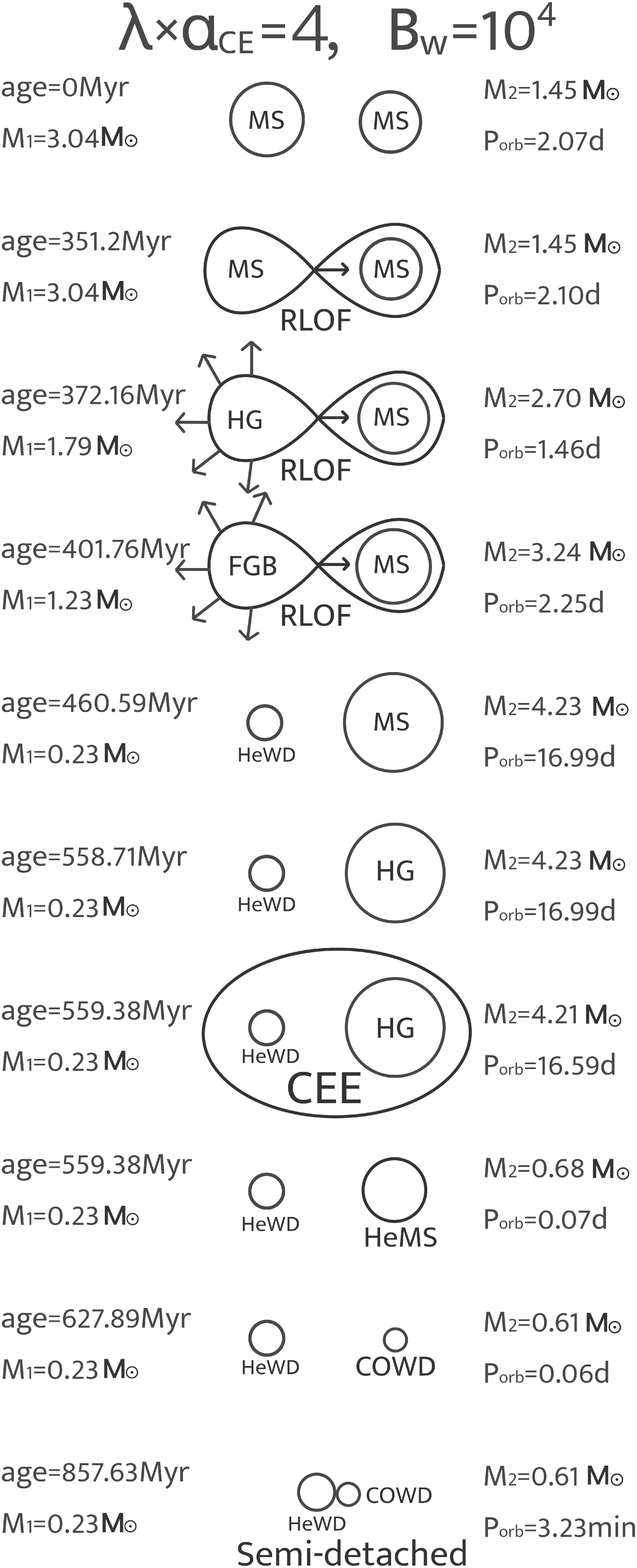}\\
\end{tabular}
\caption{Evolutionary scenarios for a binary systems with different input parameters.
The left panel represents the model with the assumptions of $\lambda\times\alpha_{\rm CE}=1.0$ and $B_{\rm W}=0$,
while the right panel for $\lambda\times\alpha_{\rm CE}=4$ and $B_{\rm W}=10^4$.
MS, HG, FGB, HeMS and RLOF mean main sequence, Hertzsprung gap, the first giant branch,
helium main sequence and Roche lobe overflow, respectively. $M_1$, $M_2$, $P_{\rm orb}$ and age
represent primary mass, secondary mass, orbital period and the evolutionary time, respectively.}
\label{fig:scenario}
\end{figure*}

In order to efficiently produce Algol systems in RS CVn binaries, \cite{Tout1988} suggested that mass-loss rate of a giant star via stellar wind
can be tidally enhanced by
\begin{equation}
\dot{M}=\dot{M}_{\rm R}[1+B_{\rm W}\max(\frac{1}{2},\frac{R}{R_{\rm L}})^6],
\end{equation}
where, $\dot{M}_{\rm R}$ is the mass-loss rate from \cite{Reimers1975}, $R$ and $R_{\rm L}$ are
the stellar and Roche-lobe radius, respectively. Here, $B_{\rm W}$ is an uncertain parameter.
Comparing with the observations, \cite{Tout1988} took $B_{\rm W}$ as $10^4$.
As Figure \ref{fig:scenario} shows, the model with $B_{\rm W}=10^4$ can
efficiently produce less massive He WD. However, this also results in a low orbital energy because the mass
lost by stellar wind carries out the orbital angular momentum. Under the assumption of $\lambda\times\alpha_{\rm CE}=$1.0,
the less massive He WD merges with its companion when this binary experiences CCE.
Therefore, we take $\lambda\times\alpha_{\rm CE}=$4 so that the binary with a low-mass He WD can survive after a CCE.
In many literatures, the parameters $\lambda$ and $\alpha_{\rm CE}$ are assumed less than 1.0.
However, $\lambda$ and $\alpha_{\rm CE}$ are very uncertain. \cite{Xu2010} calculated $\lambda$,
and found that its range changes from $\ll 0.5$ to several tens, even several hundreds.
Simultaneously, some literatures have also taken $\alpha_{\rm CE}$ larger than 1.
For example, the default value of $\alpha_{\rm CE}$ in BSE code is 3\citep{Hurley2002}.

Figure \ref{fig:scenario} gives the different products for a binary with similar initial conditions under different
assumptions: $\lambda\times\alpha_{\rm CE}=1.0$ and $B_{\rm W}=0$, or $\lambda\times\alpha_{\rm CE}=4$ and $B_{\rm W}=10^4$.
As one can see, a large $B_{\rm W}$ can produce a He WD with lower mass, and a large $\lambda\times\alpha_{\rm CE}$ can
ensure low-mass He WD survival during CEE.

\subsection{Discussions}
\begin{figure}
\includegraphics[totalheight=3.in,width=2.5in,angle=-90]{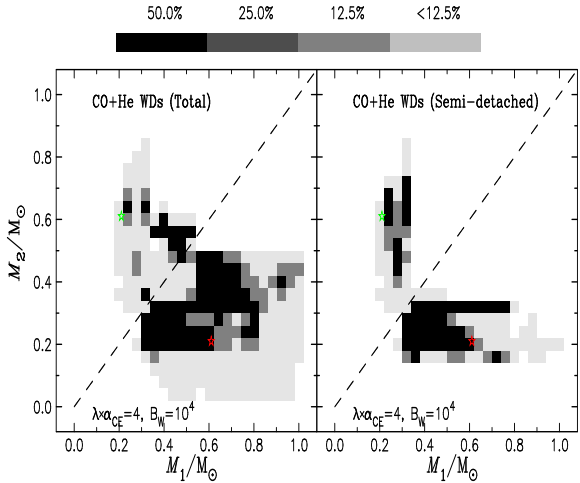}
\caption{Similar to Figure \ref{fig:m1m2c1}, but under assumptions of
$\lambda\times\alpha_{\rm CE}=4$ and $B_{\rm W}=10^4$.  }
\label{fig:m1m2c9}
\end{figure}

Figure \ref{fig:m1m2c9} gives the results of He + CO WDs via the population synthesis method
under assumptions of $\lambda\times\alpha_{\rm CE}=4$ and $B_{\rm W}=10^4$.
The primary and secondary masses in ZTFJ1539 can be covered by
He + CO WDs produced via Algol scenario. However, we must mention that
the results are sensitive to the values of two parameters.
If $B_{\rm W}<10^3$, He WD with a mass lower than about 0.27 M$_\odot$ hardly forms.
If $ \lambda \times\alpha_{\rm CE} <2$, binary systems with low-mass He WD hardly survives during CEE.

\begin{figure}
\includegraphics[totalheight=3.in,width=2.5in,angle=-90]{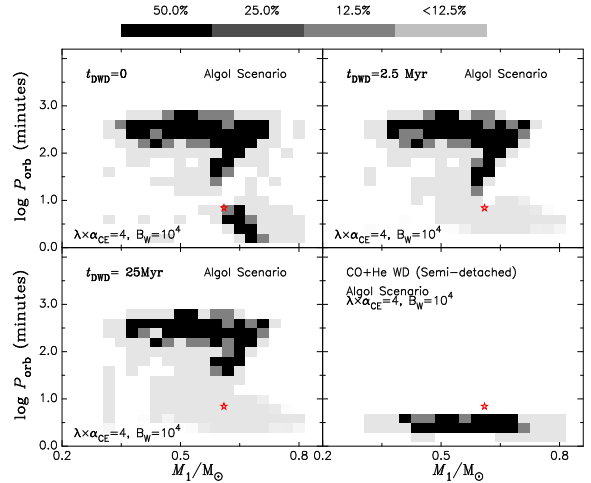}
\caption{Evolution of orbital periods of He + CO WDs produced via Algol scenario under assumptions of
$\lambda\times\alpha_{\rm CE}=4$ and $B_{\rm W}=10^4$. $t_{\rm DWD}$ represents the lifetime of DWD formation.}
\label{fig:m1poc9}
\end{figure}

Figure \ref{fig:m1poc9} shows the distributions of orbital period vs. CO WD's masse for He + CO WDs
produced via Algol scenario during different ages.
Based on the temperature of the primary WD in ZTFJ1539, \cite{Burdge2019} estimated that its cooling age should be
about 2.5 Myr. The right-top panel in Figure \ref{fig:m1poc9} gives the distribution within 2.5 Myr after
the formation of He + CO WDs. Our results cover the observations of ZTFJ1539, including the distribution
when the cooling age is about 25 Myr.
Therefore, it is possible in our binary model to produce ZTFJ1539.

In the this work, through only changing parameters $\lambda\times\alpha_{\rm CE}$ and $B_{\rm W}$,
we can explain the observational properties of ZTFJ1539 via Algol scenario.
In fact, there are still other uncertain parameters which can affect the binary evolutions, such as
mass-transfer rates and the efficiency of mass-accretion during Roche lobe overflow, rotation, the change
in orbital angular momentum caused by mass loss from the binary system, stability criteria for mass transfer, and so on.
With more DWDs like ZTFJ1539 observed in the future, it will contribute to check whether
Algol scenario can produce ZTFJ1539. Simultaneously, the observations for them
will restrain these uncertain parameters in the theoretical model for binary evolution.


\section{Conclusions}
In this work, we investigate the formation of ZTFJ1539 via nova and Algol scenarios.
Assuming that the massive WD in ZTFJ1539 just experiences a thermal runaway,
nova scenario can explain the effective temperatures of two WDs ZTFJ1539. Unfortunately,
in order to enlarging a semi-detached orbit of about 4---5 minutes to a detached orbit of about 7 minutes,
nova scenario needs a much high kick velocity of about 200 km s$^{-1}$ during nova eruption, which
can be comparable with the kick velocity of CCSN. However, the energy released by
CCSN is higher than about $10^4$ times energy from nova eruption. Simultaneously, this high kick velocity
can result in a high eccentricity of about 0.2---0.6. There is no observational evidence for
high eccentricity. Therefore, it is not a probability
to produce ZTFJ1539 by nova scenario.
If the enhancement of the mass-loss rate of a giant star via
stellar wind triggered by tidal effect and a high efficiency of ejecting CE are considered,
Algol scenario can also produce ZTFJ1539.

As a significant source of gravitational radiation close to the peak of LISA¡¯s sensitivity,
ZTFJ1539 has attracted attention. In the meantime, ZTFJ1539's progenitor and its destination
involve many unsolved physical processes of binary evolutions, such as CEE, mass loss, and binary merger, and so on.
It is worth investigating ZTFJ1539 further.

\section*{Acknowledgments}
This work received the generous support of the National Natural Science Foundation of China,
project Nos. 11863005, 11763007, 11803026, 11563008£¬11473024, 11463005, and 11503008.
We would also like to express our gratitude to the Tianshan Youth Project of Xinjiang No. 2017Q014.

\bibliography{lglmn}

\label{lastpage}

\end{document}